# COMPENDIUM OF EIGENMODES IN THIRD HARMONIC CAVITIES FOR FLASH AND THE XFEL

**Ian Reginald Roy Shinton and Nawin Juntong; Cockcroft Inst./University of Manchester, U.K.**

**Abstract**
The resonant modes in the 9cell 3.9GHz bunch shaping cavity designed by FERMILAB in collaboration with DESY [1] and installed in FLASH at DESY were calculated up to the range of 10GHz in terms of the band structure of this design. The modal nature of this structure has previously been investigated by various parties [1]. We have extended this work to include a modal pictorial dictionary in which the nature of the modes can be readily identified as well as the R/Q's for each of the modes. Below 10GHz only monopole, dipole, quadrupole and sextupole bands exist for this particular structure. Herein we only consider the modal patterns of the bands themselves and have not included the beampipe modes in the pictorial dictionary. The R/Q definition that we use is that of [2]. In addition to the finite element simulations we also utilise a capacitive-inductive circuit model to achieve a rapid characterisation of the cavity.

**Contents**

# Electric field boundary conditions applied to beam-pipe boundaries

## Monopole modes

These monopole results were generated using HFSSv11 in which only a 10 degree slice of the structure was simulated using : magnetic (M) symmetry planes. Electric (E) boundary conditions were applied to the ends of the beampipes. The meshing process consisted of: a volume mesh of applying a volumetric mesh restricted to a cell length no greater that 5mm; a similar mesh was applied on the surface of the structure in which the maximum mesh cell length was restricted to 5mm; in order to better account for the curved nature of the surface geometry a segmentation of 10 slices was applied. All simulations were run until a convergence of less than 0.05% was achieved. Both the electric and magnetic energies were used in the R/Q calculations

| | $\omega/2\pi$ (GHz) | Band type | R/Q: $\Omega$ |
|---|---|---|---|
| 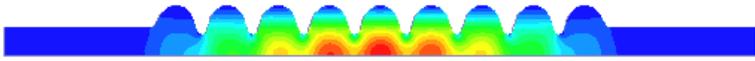 | 3.7483 | M Band 1 #1 boundary EE | 0.009 |
| 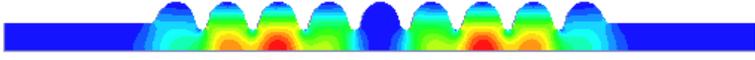 | 3.7617 | M Band 1 #2 boundary EE | 0.054 |
| 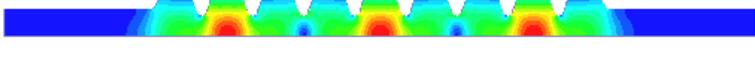 | 3.7825 | M Band 1 #3 boundary EE | 0.075 |
| 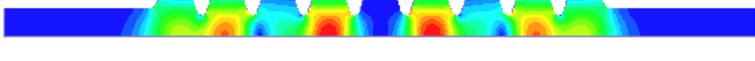 | 3.8081 | M Band 1 #4 boundary EE | 0.160 |
| 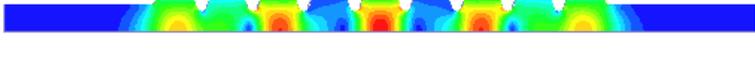 | 3.8357 | M Band 1 #5 boundary EE | 0.294 |
| 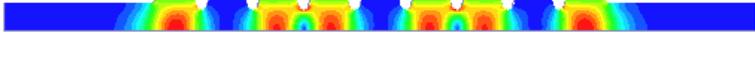 | 3.8617 | M Band 1 #6 boundary EE | 0.265 |
| 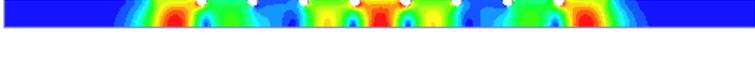 | 3.8832 | M Band 1 #7 boundary EE | 0.153 |
| 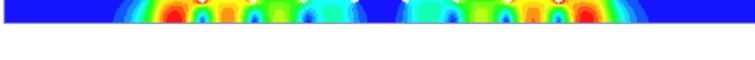 | 3.8972 | M Band 1 #8 boundary EE | 3.303 |
| 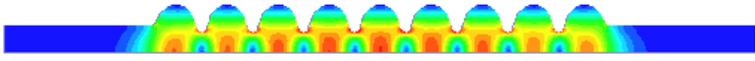 | 3.9028 | M Band 1 #9 boundary EE | 372.705 |
| 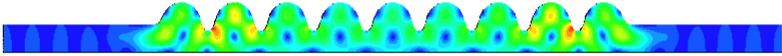 | 7.039 | M Band 2 #1 boundary EE | 0.05 |
| 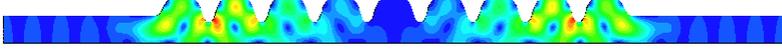 | 7.0685 | M Band 2 #2 boundary EE | 0.30 |
| 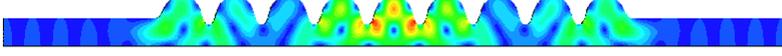 | 7.1088 | M Band 2 #3 boundary EE | 3.15 |
| 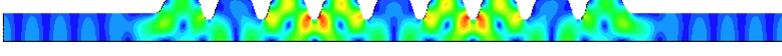 | 7.1535 | M Band 2 #4 boundary EE | 1.91 |
| 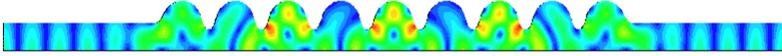 | 7.2006 | M Band 2 #5 boundary EE | 11.11 |
| 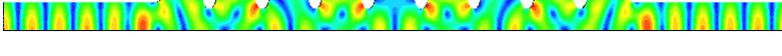 | 7.2581 | M Band 2 #6 boundary EE | 35.10 |
| 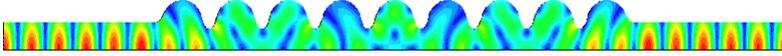 | 7.3292 | M Band 2 #7 boundary EE | 0.523 |
| 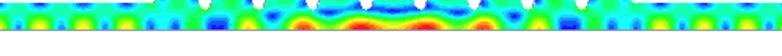 | 7.5091 | M Band 2 #8 boundary EE | 19.122 |
| 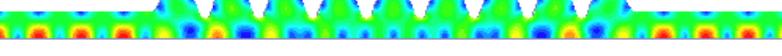 | 7.5829 | M Band 2 #9 boundary EE | 47.534 |
| | | | |
| 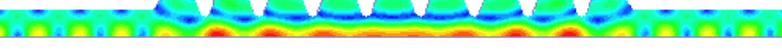 | 7.6417 | M Band 3 #1 boundary EE | 0.594 |
| 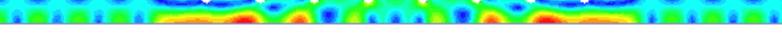 | 7.7230 | M Band 3 #2 boundary EE | 21.894 |
| 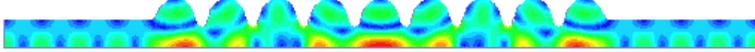 | 7.8027 | M Band 3 #3 boundary EE | 0.229 |
| 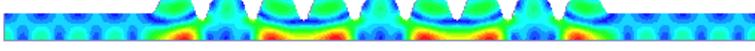 | 7.8798 | M Band 3 #4 boundary EE | 1.850 |
| 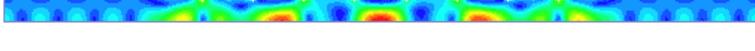 | 7.9536 | M Band 3 #5 boundary EE | 0.639 |
| 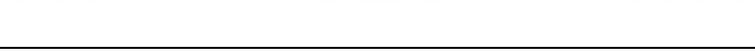 | 8.0215 | M Band 3 #6 boundary EE | 0.015 |
| 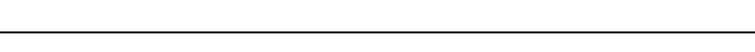 | 8.0815 | M Band 3 #7 boundary EE | 0.113 |
| 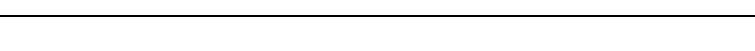 | 8.1295 | M Band 3 #8 boundary EE | 0.010 |
| 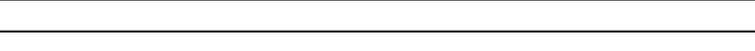 | 8.1612 | M Band 3 #9 boundary EE | 0.002 |
| 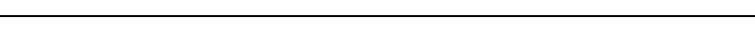 | 9.7927 | M Band 4 #1 boundary EE | 0.000 |
| 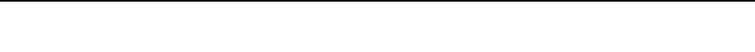 | 9.8308 | M Band 4 #2 boundary EE | 1.457 |
| 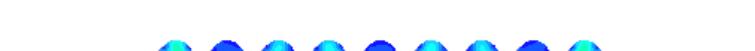 | 9.8840 | M Band 4 #3 boundary EE | 0.374 |
| 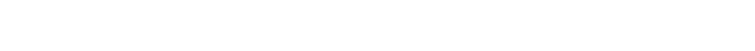 | 9.9368 | M Band 4 #4 boundary EE | 7.535 |
| 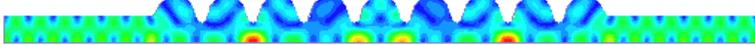 | 9.9865 | M Band 4 #5 boundary EE | 0.010 |
| 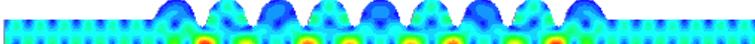 | 10.0581 | M Band 4 #6 boundary EE | 0.862 |
| 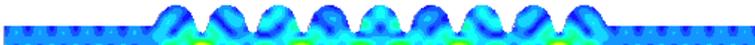 | 10.1645 | M Band 4 #7 boundary EE | 1.194 |
| 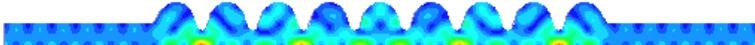 | 10.2960 | M Band 4 #8 boundary EE | 0.760 |
| 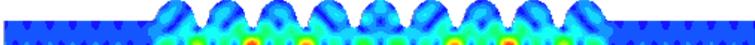 | 10.4418 | M Band 4 #9 boundary EE | 0.243 |

**Monopole cut-off parameters**

$TM_{01}$: $f_c$ = 5.7375 GHz for iris radius 15 mm
   $f_c$ = 7.6501 GHz for beam pipe radius 20 mm

$TE_{11}$: $f_c$ = 4.3920 GHz for iris radius 15 mm
   $f_c$ = 5.8560 GHz for beam pipe radius 20 mm

## Dipole modes

These dipole results were generated using HFSSv11 in which only a 90 degree slice of the structure was simulated using both magnetic (M) and electric (E) symmetry planes. Electric (E) boundary conditions were applied to the ends of the beampipes. The meshing process consisted of: a volume mesh of applying a volumetric mesh restricted to a cell length no greater that 5mm; a similar mesh was applied on the surface of the structure in which the maximum mesh cell length was restricted to 5mm; in order to better account for the curved nature of the surface geometry a segmentation of 20 slices was applied. All simulations were run until a convergence of less than 0.05% was achieved. In the R/Q calculation in order to reduce the computational time required we considered only the electric energy in the calculation.

| Electric field pattern | $\omega/2\pi$ (GHz) | Band type | R/Q: $\Omega/cm^2$ |
|---|---|---|---|
| 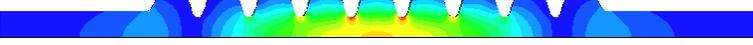 | 4.2953 | D Band 1 #1 boundary EE | 0.00 |
| 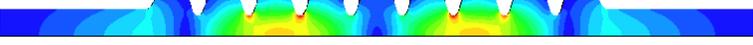 | 4.3580 | D Band 1 #2 boundary EE | 0.29 |
| 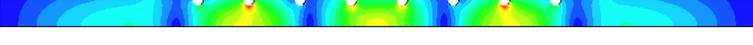 | 4.4460 | D Band 1 #3 boundary EE | 0.00 |
| 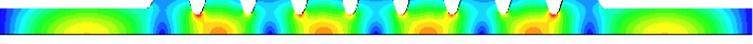 | 4.5388 | D Band 1 #4 boundary EE | 1.08 |
| 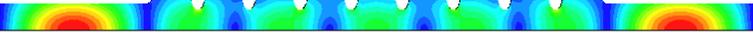 | 4.5972 | D Band 1 #5 boundary EE | 0.79 |
| 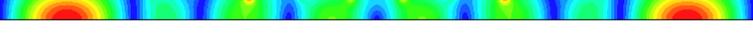 | 4.6399 | D Band 1 #6 boundary EE | 0.16 |
| 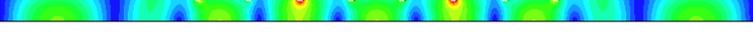 | 4.7227 | D Band 1 #7 boundary EE | 10.37 |
| 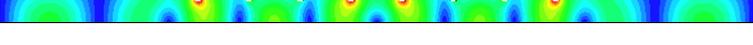 | 4.8312 | D Band 1 #8 boundary EE | 50.20 |
| 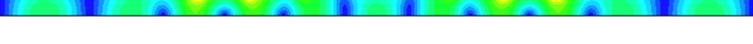 | 4.9260 | D Band 1 #9 boundary EE | 30.38 |
| 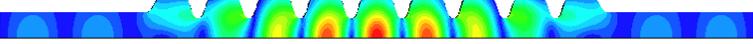 | 5.3583 | D Band 2 #1 boundary EE | 0.04 |
| 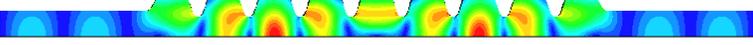 | 5.4058 | D Band 2 #2 boundary EE | 5.01 |
| 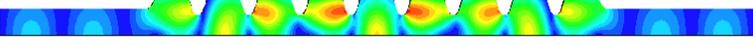 | 5.4441 | D Band 2 #3 boundary EE | 20.88 |
| 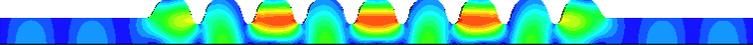 | 5.4696 | D Band 2 #4 boundary EE | 16.07 |
| 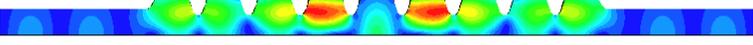 | 5.4849 | D Band 2 #5 boundary EE | 0.98 |
| 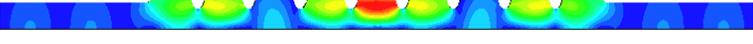 | 5.4933 | D Band 2 #6 boundary EE | 1.25 |
| 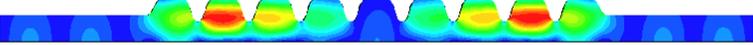 | 5.4973 | D Band 2 #7 boundary EE | 0.31 |
| 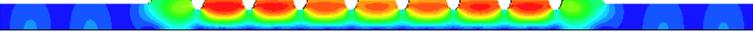 | 5.4982 | D Band 2 #8 boundary EE | 0.48 |
| 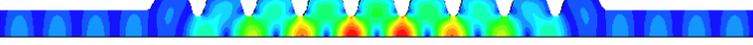 | 6.8225 | D Band 3 #1 boundary EE | 0.01 |
| 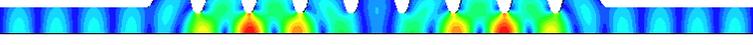 | 6.8994 | D Band 3 #2 boundary EE | 0.03 |
| 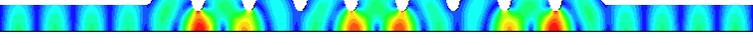 | 7.0022 | D Band 3 #3 boundary EE | 0.06 |
| 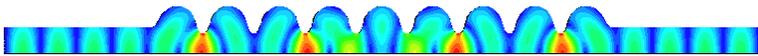 | 7.1227 | D Band 3 #4 boundary EE | 0.18 |
| 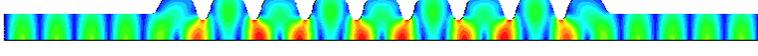 | 7.2551 | D Band 3 #5 boundary EE | 0.55 |
| 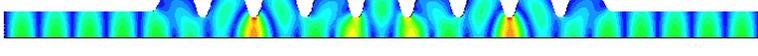 | 7.3852 | D Band 3 #6 boundary EE | 0.01 |
| 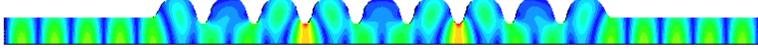 | 7.4915 | D Band 3 #7 boundary EE | 0.45 |
| 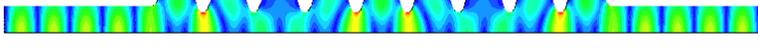 | 7.5654 | D Band 3 #8 boundary EE | 0.27 |
| 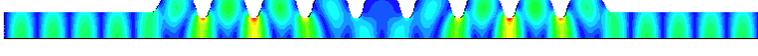 | 7.6235 | D Band 3 #9 boundary EE | 1.27 |
| 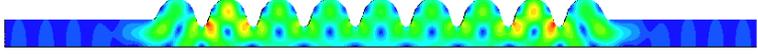 | 8.4995 | D Band 4 #1 boundary EE | 0.13 |
| 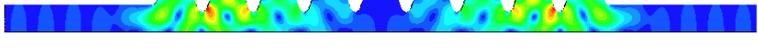 | 8.5045 | D Band 4 #2 boundary EE | 0.09 |
| 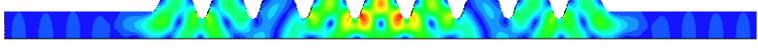 | 8.5318 | D Band 4 #3 boundary EE | 0.16 |
| 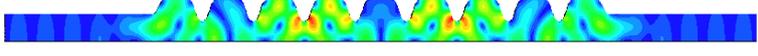 | 8.5762 | D Band 4 #4 boundary EE | 0.11 |
| 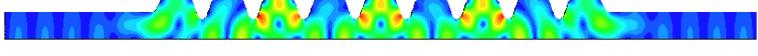 | 8.6400 | D Band 4 #5 boundary EE | 0.42 |
| 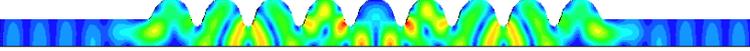 | 8.7207 | D Band 4 #6 boundary EE | 1.04 |
| 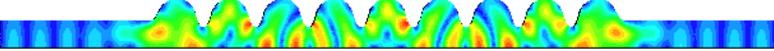 | 8.7998 | D Band 4 #7 boundary EE | 10.06 |
| 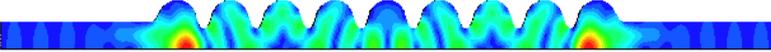 | 8.8613 | D Band 4 #8 boundary EE | 2.57 |
| 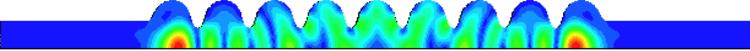 | 8.9167 | D Band 4 #9 boundary EE | 0.39 |
| 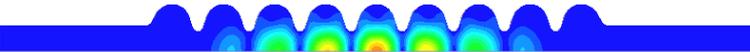 | 9.0560 | D Band 5 #1 boundary EE | 0.00 |
| 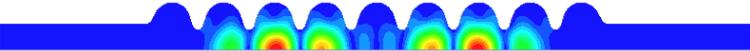 | 9.0568 | D Band 5 #2 boundary EE | 0.05 |
| 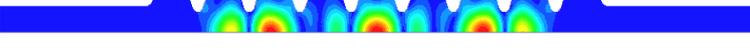 | 9.0585 | D Band 5 #3 boundary EE | 0.07 |
| 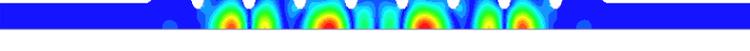 | 9.0620 | D Band 5 #4 boundary EE | 2.17 |
| 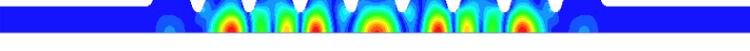 | 9.0703 | D Band 5 #5 boundary EE | 4.04 |
| 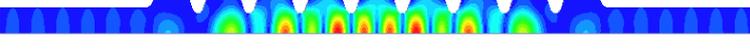 | 9.0933 | D Band 5 #6 boundary EE | 0.55 |
| 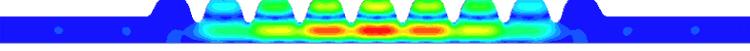 | 9.6937 | D Band 6 #1 boundary EE | 0.00 |
| 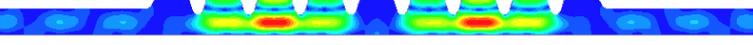 | 9.7117 | D Band 6 #2 boundary EE | 0.01 |
| 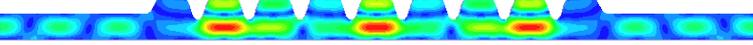 | 9.7393 | D Band 6 #3 boundary EE | 0.13 |
| 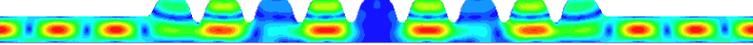 | 9.7665 | D Band 6 #4 boundary EE | 0.34 |
| 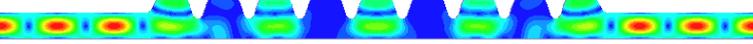 | 9.7829 | D Band 6 #5 boundary EE | 5.92 |
| 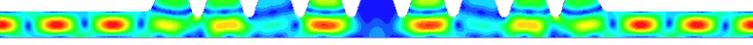 | 9.7977 | D Band 6 #6 boundary EE | 0.78 |
| 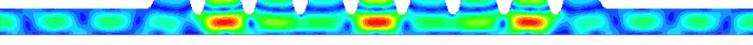 | 9.8243 | D Band 6 #7 boundary EE | 0.09 |
| 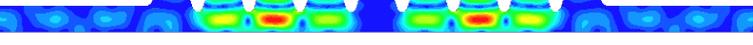 | 9.8520 | D Band 6 #8 boundary EE | 0.01 |
| 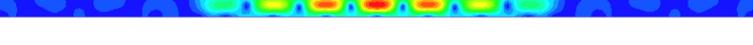 | 9.8722 | D Band 6 #9 boundary EE | 0.00 |

**Dipole cut-off parameters**

$TM_{01}$: $f_c$ = 5.7375 GHz for iris radius 15 mm
$f_c$ = 7.6501 GHz for beam pipe radius 20 mm

$TE_{11}$: $f_c$ = 4.3920 GHz for iris radius 15 mm
$f_c$ = 5.8560 GHz for beam pipe radius 20 mm

## Quadrupole modes

These quadrupole results were generated using HFSSv11 in which only a 90 degree slice of the structure was simulated using magnetic (M) symmetry planes. Electric (E) boundary conditions were applied to the ends of the beampipes. The meshing process consisted of: a volume mesh of applying a volumetric mesh restricted to a cell length no greater that 5mm; a similar mesh was applied on the surface of the structure in which the maximum mesh cell length was restricted to 5mm; in order to better account for the curved nature of the surface geometry a segmentation of 20 slices was applied. All simulations were run until a convergence of less than 0.05% was achieved. In the R/Q calculation in order to reduce the computational time required we considered only the electric energy in the calculation.

| Electric field pattern | $\omega/2\pi$ (GHz) | Band type | R/Q: $\Omega/cm^4$ |
|---|---|---|---|
| 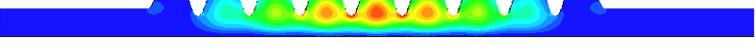 | 6.5625 | Q Band 1 #1 boundary EE | 0.18 |
| 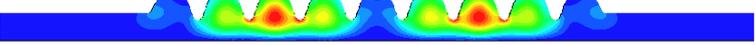 | 6.5832 | Q Band 1 #2 boundary EE | 3.75 |
| 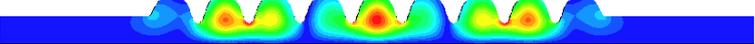 | 6.6158 | Q Band 1 #3 boundary EE | 4.38 |
| 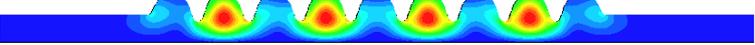 | 6.6578 | Q Band 1 #4 boundary EE | 0.19 |
| 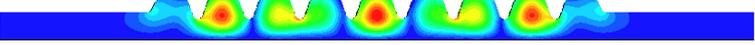 | 6.7057 | Q Band 1 #5 boundary EE | 0.31 |
| 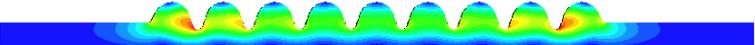 | 6.9977 | Q Band 2 #1 boundary EE | 0.14 |
| 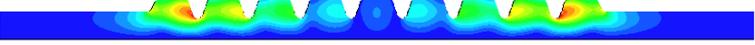 | 7.0067 | Q Band 2 #2 boundary EE | 0.08 |
| 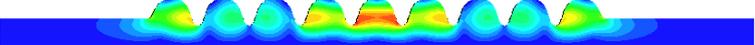 | 7.0429 | Q Band 2 #3 boundary EE | 0.15 |
| 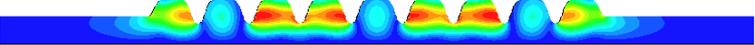 | 7.0798 | Q Band 2 #4 boundary EE | 0.00 |
| 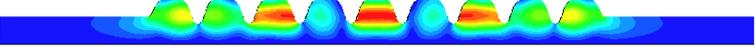 | 7.1133 | Q Band 2 #5 boundary EE | 0.22 |
| 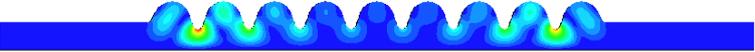 | 9.1103 | Q Band 3 #1 boundary EE | 1.86 |
| 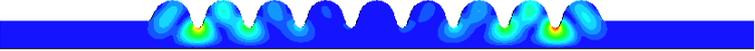 | 9.1106 | Q Band 3 #2 boundary EE | 0.01 |
| 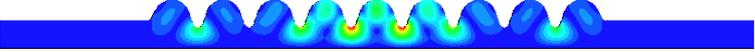 | 9.1204 | Q Band 3 #3 boundary EE | 2.39 |
| 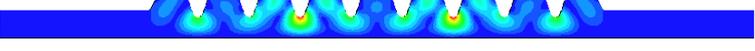 | 9.1314 | Q Band 3 #4 boundary EE | 4.74 |
| 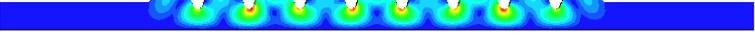 | 9.1468 | Q Band 3 #5 boundary EE | 2.26 |
| 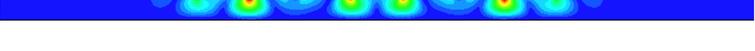 | 9.1656 | Q Band 3 #6 boundary EE | 0.00 |
| 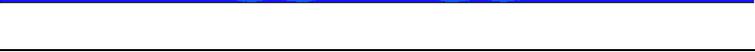 | 9.1853 | Q Band 3 #7 boundary EE | 0.08 |

**Quadrupole cut-off parameters**

$TM_{01}$: $f_c$ = 5.7375 GHz for iris radius 15 mm
$f_c$ = 7.6501 GHz for beam pipe radius 20 mm

$TE_{11}$: $f_c$ = 4.3920 GHz for iris radius 15 mm
$f_c$ = 5.8560 GHz for beam pipe radius 20 mm

## Sextupole modes

These sextupole results were generated using HFSSv11 in which only a 90 degree slice of the structure was simulated using both magnetic (M) and electric (E) symmetry planes. Electric (E) boundary conditions were applied to the ends of the beampipes. The meshing process consisted of: a volume mesh of applying a volumetric mesh restricted to a cell length no greater that 5mm; a similar mesh was applied on the surface of the structure in which the maximum mesh cell length was restricted to 5mm; in order to better account for the curved nature of the surface geometry a segmentation of 20 slices was applied. All simulations were run until a convergence of less than 0.05% was achieved. In the R/Q calculation in order to reduce the computational time required we considered only the electric energy in the calculation.

Here we have only considered the first sextupole band as we were interested in the band structure up to 10GHz, the majority of the second sextupole band lies above this frequency – refer to dispersion curve for a middle cell structure in the Dispersion curve section.

| Electric field pattern | $\omega/2\pi$ (GHz) | Band type | R/Q: $\Omega/cm^6$ |
|---|---|---|---|
| 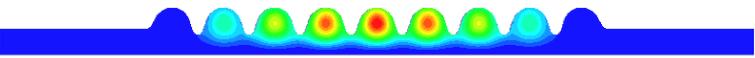 | 8.1883 | S Band 1 #1 boundary EE | 0.51 |
| 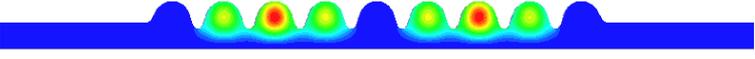 | 8.1930 | S Band 1 #2 boundary EE | 0.20 |
| 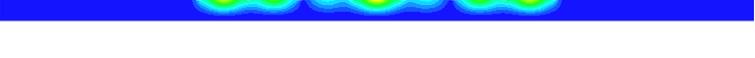 | 8.2001 | S Band 1 #3 boundary EE | 0.01 |
| 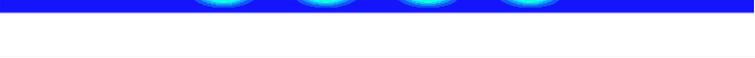 | 8.2086 | S Band 1 #4 boundary EE | 0.03 |
| 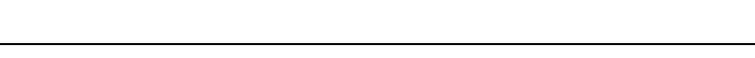 | 8.2174 | S Band 1 #5 boundary EE | 0.00 |
| 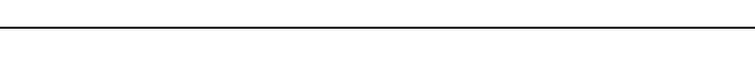 | 8.2250 | S Band 1 #6 boundary EE | 0.00 |
| 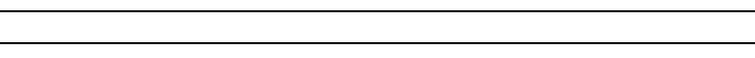 | 8.2301 | S Band 1 #7 boundary EE | 0.00 |
| 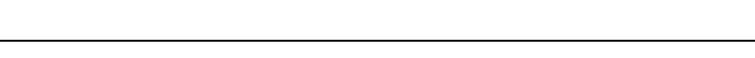 | 8.8020 | S Band 2 #1 boundary EE | 0.06 |
| 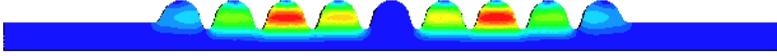 | 8.8086 | S Band 2 #2 boundary EE | 0.00 |
| 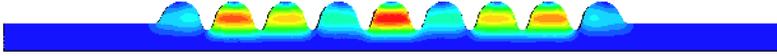 | 8.8179 | S Band 2 #3 boundary EE | 0.00 |
| 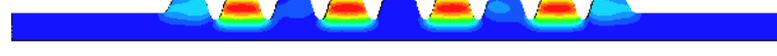 | 8.8275 | S Band 2 #4 boundary EE | 0.00 |
| 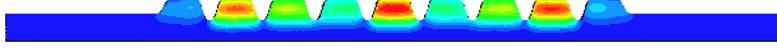 | 8.8375 | S Band 2 #5 boundary EE | 0.03 |
| 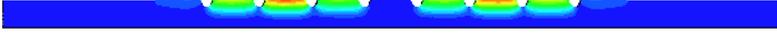 | 8.8452 | S Band 2 #6 boundary EE | 0.04 |
| 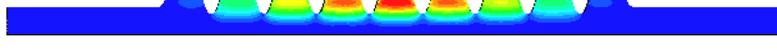 | 8.8503 | S Band 1 #7 boundary EE | 0.01 |

**Sextupole cut-off parameters**

$TM_{01}$: $f_c$ = 5.7375 GHz for iris radius 15 mm
$f_c$ = 7.6501 GHz for iris radius 20 mm
$TE_{11}$: $f_c$ = 4.3920 GHz for iris radius 15 mm
$f_c$ = 5.8560 GHz for iris radius 20 mm

# Comparison to MAFIA results

In the table below a direct comparison to the MAFIA modal results presented in [1] is given. In general the results from both codes are consistent. Note although this work is inclusive of the HOM bands below 10GHz, the MAFIA comparison of [1] did not include all the bands up to this frequency; consequently only direct comparison to the results in the literature are displayed herein and as such there is no MAFIA sextupole comparison presented here.

**Monopole**

| Band | HFSS | | | MAFIA | |
|---|---|---|---|---|---|
| | f: GHz | R/Q: Ω | | f: GHz | R/Q: Ω |
| 1 | 3.7483 | 0.009 | | 3.7455 | 0.007 |
| 1 | 3.7617 | 0.054 | | 3.7589 | 0.06 |
| 1 | 3.7825 | 0.075 | | 3.7796 | 0.085 |
| 1 | 3.8081 | 0.160 | | 3.8051 | 0.166 |
| 1 | 3.8357 | 0.294 | | 3.8325 | 0.278 |
| 1 | 3.8617 | 0.265 | | 3.8585 | 0.194 |
| 1 | 3.8832 | 0.153 | | 3.8799 | 0.317 |
| 1 | 3.8972 | 3.303 | | 3.894 | 0.134 |
| 1 | 3.9028 | 372.705 | | 3.8991 | 375.748 |
| | | | | | |
| 2 | 7.043494 | 0.142 | | 7.041 | 0.225 |
| 2 | 7.081938 | 0.976 | | 7.079 | 1.5886 |
| 2 | 7.138101 | 0.093 | | 7.1349 | 0.123 |
| 2 | 7.207645 | 2.924 | | 7.2036 | 4.334 |
| 2 | 7.283561 | 0.645 | | 7.2799 | 1.204 |
| 2 | 7.362203 | 2.673 | | 7.3585 | 3.594 |
| 2 | 7.437698 | 10.705 | | 7.4344 | 15.939 |
| 2 | 7.509145 | 19.122 | | 7.5054 | 23.219 |
| 2 | 7.582914 | 47.534 | | 7.5765 | 42.874 |
| | | | | | |
| 3 | 7.641745 | 0.594 | | 7.6398 | 2.134 |
| 3 | 7.723047 | 21.894 | | 7.718 | 18.59 |
| 3 | 7.802748 | 0.229 | | 7.7964 | 0.621 |
| 3 | 7.879784 | 1.850 | | 7.8737 | 0.876 |
| 3 | 7.953646 | 0.639 | | 7.9477 | 0.576 |
| 3 | 8.021515 | 0.015 | | 8.0163 | 0.019 |
| 3 | 8.081535 | 0.113 | | 8.0768 | 0.021 |
| 3 | 8.129549 | 0.010 | | 8.12255 | 0.04 |
| 3 | 8.161176 | 0.002 | | 8.1578 | 0.005 |

**Dipole**

| Band | HFSS | | | MAFIA | |
|---|---|---|---|---|---|
| | f: GHz | R/Q:$\Omega$/cm$^2$ | | f: GHz | R/Q:$\Omega$/cm$^2$ |
| 1 | 4.2953 | 0.00 | | 4.3019 | 0.00 |
| 1 | 4.3580 | 0.29 | | 4.3641 | 0.29 |
| 1 | 4.4460 | 0.00 | | 4.4514 | 0.00 |
| 1 | 4.5388 | 1.08 | | 4.5428 | 1.07 |
| 1 | 4.5972 | 0.79 | | 4.5993 | 0.82 |
| 1 | 4.6399 | 0.16 | | 4.6422 | 0.13 |
| 1 | 4.7227 | 10.37 | | 4.726 | 10.39 |
| 1 | 4.8312 | 50.20 | | 4.8341 | 50.70 |
| 1 | 4.9260 | 30.38 | | 4.9282 | 30.41 |
| | | | | | |
| 2 | 5.3583 | 0.04 | | 5.3588 | 0.04 |
| 2 | 5.4058 | 5.01 | | 5.4052 | 4.99 |
| 2 | 5.4441 | 20.88 | | 5.4427 | 20.91 |
| 2 | 5.4696 | 16.07 | | 5.4676 | 16.01 |
| 2 | 5.4849 | 0.98 | | 5.4826 | 0.96 |
| 2 | 5.4933 | 1.25 | | 5.4908 | 1.25 |
| 2 | 5.4973 | 0.31 | | 5.4946 | 0.31 |
| 2 | 5.4982 | 0.48 | | 5.4955 | 0.47 |
| | | | | | |
| 3 | 6.8225 | 0.01 | | 6.8264 | 0.01 |
| 3 | 6.8994 | 0.03 | | 6.9027 | 0.04 |
| 3 | 7.0022 | 0.06 | | 7.005 | 0.06 |
| 3 | 7.1227 | 0.18 | | 7.1248 | 0.19 |
| 3 | 7.2551 | 0.55 | | 7.2563 | 0.57 |
| 3 | 7.3852 | 0.01 | | 7.3853 | 0.01 |
| 3 | 7.4915 | 0.45 | | 7.4907 | 0.47 |
| 3 | 7.5654 | 0.27 | | 7.5636 | 0.26 |
| 3 | 7.6235 | 1.27 | | 7.6206 | 1.37 |
| | | | | | |
| 4 | 8.4995 | 0.13 | | 8.4912 | 0.11 |
| 4 | 8.5045 | 0.09 | | 8.4968 | 0.09 |
| 4 | 8.5318 | 0.16 | | 8.524 | 0.15 |
| 4 | 8.5762 | 0.11 | | 8.569 | 0.11 |
| 4 | 8.6399 | 0.42 | | 8.6336 | 0.4 |
| 4 | 8.7207 | 1.04 | | 8.7158 | 1.03 |
| 4 | 8.7998 | 10.06 | | 8.8001 | 10.35 |
| 4 | 8.8613 | 2.57 | | 8.863 | 2.59 |
| 4 | 8.9167 | 0.39 | | 8.9185 | 0.27 |
| | | | | | |
| 5 | 9.0560 | 0.00 | | 9.0594 | 0.00 |
| 5 | 9.0568 | 0.05 | | 9.0602 | 0.05 |
| 5 | 9.0585 | 0.07 | | 9.0618 | 0.06 |
| 5 | 9.0620 | 2.17 | | 9.0653 | 2.19 |
| 5 | 9.0703 | 4.04 | | 9.0736 | 4.06 |
| 5 | 9.0933 | 0.55 | | 9.0961 | 0.56 |

**Quadrupole**

| Band | HFSS | | | MAFIA | |
|---|---|---|---|---|---|
| | f: GHz | R/Q:$\Omega$/cm$^4$ | | f: GHz | R/Q:$\Omega$/cm$^4$ |
| 1 | 6.5625 | 0.18 | | 6.5694 | 0.19 |
| 1 | 6.5832 | 3.75 | | 6.5895 | 3.78 |
| 1 | 6.6158 | 4.38 | | 6.6221 | 4.34 |
| 1 | 6.6578 | 0.19 | | 6.6627 | 0.16 |
| 1 | 6.7057 | 0.31 | | 6.7093 | 0.30 |
| | | | | | |
| 2 | 6.9977 | 0.14 | | 6.997 | 0.13 |
| 2 | 7.0067 | 0.08 | | 7.0056 | 0.08 |
| 2 | 7.0429 | 0.15 | | 7.042 | 0.15 |
| 2 | 7.0798 | 0.00 | | 7.0773 | 0.00 |
| 2 | 7.1133 | 0.22 | | 7.1106 | 0.25 |
| | | | | | |
| 3 | 9.1103 | 1.86 | | 9.1106 | 3.26 |
| 3 | 9.1106 | 0.01 | | 9.1115 | 0.16 |
| 3 | 9.1204 | 2.39 | | 9.1198 | 7.70 |
| 3 | 9.1314 | 4.74 | | 9.1323 | 11.10 |
| 3 | 9.1468 | 2.26 | | 9.1513 | 2.80 |
| 3 | 9.1656 | 0.00 | | 9.1738 | 0.03 |
| 3 | 9.1853 | 0.08 | | 9.196 | 0.22 |

# Magnetic field boundary conditions applied to beam-pipe boundaries

### **Monopole modes**

These monopole results were generated using HFSSv11 in which only a 10 degree slice of the structure was simulated using magnetic (M) symmetry planes. Magnetic (M) boundary conditions were applied to the ends of the beampipes. The meshing process consisted of: a volume mesh of applying a volumetric mesh restricted to a cell length no greater that 5mm; a similar mesh was applied on the surface of the structure in which the maximum mesh cell length was restricted to 5mm; in order to better account for the curved nature of the surface geometry a segmentation of 10 slices was applied. All simulations were run until a convergence of less than 0.05% was achieved. Both the electric and magnetic energies were used in the R/Q calculations

| | $\omega/2\pi$ (GHz) | Band type | R/Q: $\Omega$ |
|---|---|---|---|
| 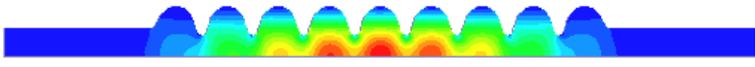 | 3.7483 | M Band 1 #1 boundary MM | 0.009 |
| 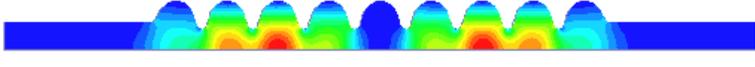 | 3.7617 | M Band 1 #2 boundary MM | 0.054 |
| 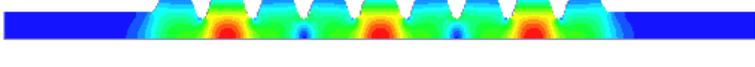 | 3.7825 | M Band 1 #3 boundary MM | 0.075 |
| 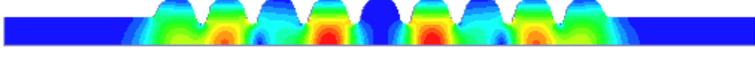 | 3.8081 | M Band 1 #4 boundary MM | 0.160 |
| 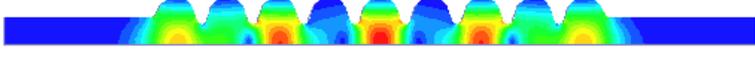 | 3.8357 | M Band 1 #5 boundary MM | 0.294 |
| 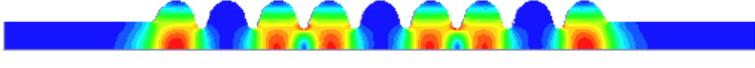 | 3.8617 | M Band 1 #6 boundary MM | 0.265 |
| 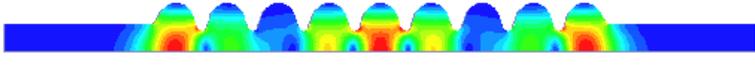 | 3.8832 | M Band 1 #7 boundary MM | 0.153 |
| 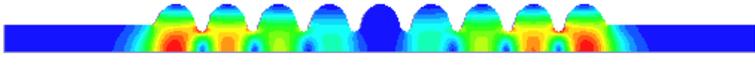 | 3.8972 | M Band 1 #8 boundary MM | 3.303 |
| 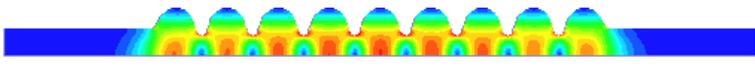 | 3.9028 | M Band 1 #9 boundary MM | 372.705 |
| 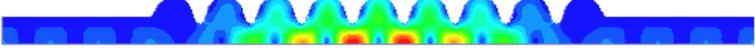 | 7.0391 | M Band 2 #1 boundary MM | 0.006 |
| 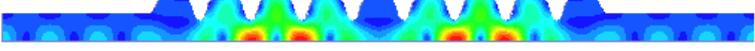 | 7.0685 | M Band 2 #2 boundary MM | 0.0338 |
| 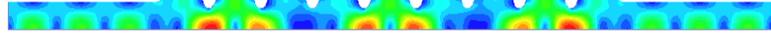 | 7.1087 | M Band 2 #3 boundary MM | 0.350 |
| 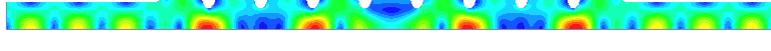 | 7.1535 | M Band 2 #4 boundary MM | 0.212 |
| 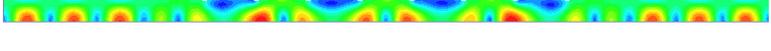 | 7.2006 | M Band 2 #5 boundary MM | 1.235 |
| 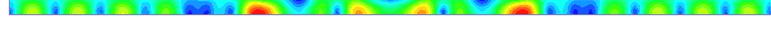 | 7.2581 | M Band 2 #6 boundary MM | 3.400 |
| 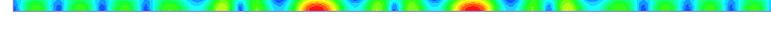 | 7.3292 | M Band 2 #7 boundary MM | 0.058 |
| 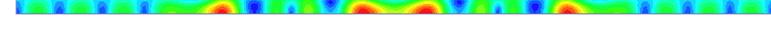 | 7.4105 | M Band 2 #8 boundary MM | 15.076 |
| 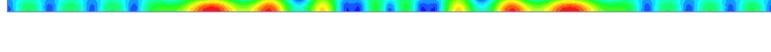 | 7.4957 | M Band 2 #9 boundary MM | 28.128 |
| 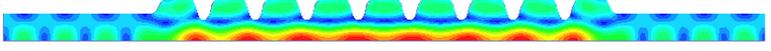 | 7.5768 | M Band 3 #1 boundary MM | 3.604 |
| 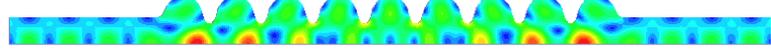 | 7.6760 | M Band 3 #2 boundary MM | 55.521 |
| 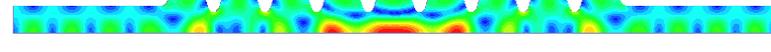 | 7.7562 | M Band 3 #3 boundary MM | 0.109 |
| | 7.8315 | M Band 3 #4 boundary MM | 0.735 |
| | 7.9007 | M Band 3 #5 boundary MM | 0.004 |
| | 7.9616 | M Band 3 #6 boundary MM | 0.259 |
| | 8.0142 | M Band 3 #7 boundary MM | 0.693 |
| | 8.0603 | M Band 3 #8 boundary MM | 0.115 |
| | 8.1023 | M Band 3 #9 boundary MM | 0.731 |
| | 9.7942 | M Band 4 #1 boundary MM | 0.000 |
| | 9.8423 | M Band 4 #2 boundary MM | 0.071 |
| | 9.9156 | M Band 4 #3 boundary MM | 0.421 |
| | 10.0128 | M Band 4 #4 boundary MM | 5.164 |
| | 10.1292 | M Band 4 #5 boundary MM | 0.274 |
| | 10.2571 | M Band 4 #6 boundary MM | 0.6894 |
| 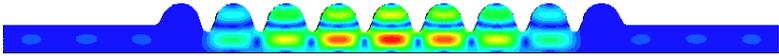 | 10.3728 | M Band 4 #7 boundary MM | 2.159 |
| 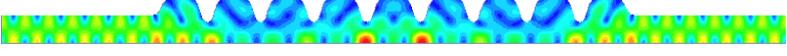 | 10.4445 | M Band 4 #8 boundary MM | 0.330 |
| 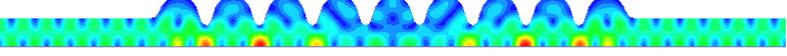 | 10.5081 | M Band 4 #9 boundary MM | 0.750 |

**Monopole cut-off parameters**

$TM_{01}$: $f_c$ = 5.7375 GHz for iris radius 15 mm
    $f_c$ = 7.6501 GHz for beam pipe radius 20 mm

$TE_{11}$: $f_c$ = 4.3920 GHz for iris radius 15 mm
    $f_c$ = 5.8560 GHz for beam pipe radius 20 mm

## Dipole modes

These dipole results were generated using HFSSv11 in which only a 90 degree slice of the structure was simulated using both magnetic (M) and electric (E) symmetry planes. Magnetic (M) boundary conditions were applied to the ends of the beampipes. The meshing process consisted of: a volume mesh of applying a volumetric mesh restricted to a cell length no greater that 5mm; a similar mesh was applied on the surface of the structure in which the maximum mesh cell length was restricted to 5mm; in order to better account for the curved nature of the surface geometry a segmentation of 20 slices was applied. All simulations were run until a convergence of less than 0.05% was achieved. In the R/Q calculation in order to reduce the computational time required we considered only the electric energy in the calculation.

| Electric field pattern | $\omega/2\pi$ (GHz) | Band type | R/Q: $\Omega/cm^2$ |
|---|---|---|---|
| 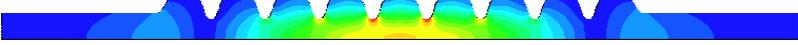 | 4.2951 | D Band 1 #1 boundary MM | 0.00 |
| 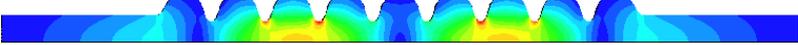 | 4.3566 | D Band 1 #2 boundary MM | 0.26 |
| 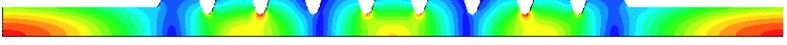 | 4.4285 | D Band 1 #3 boundary MM | 0.07 |
| 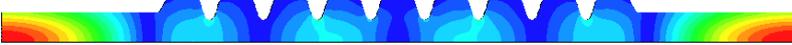 | 4.4498 | D Band 1 #4 boundary MM | 0.00 |
| 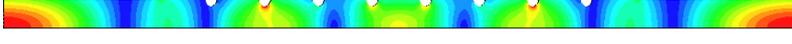 | 4.4751 | D Band 1 #5 boundary MM | 0.32 |
| 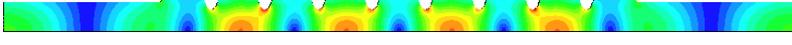 | 4.5683 | D Band 1 #6 boundary MM | 1.21 |
| 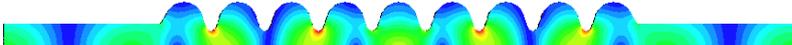 | 4.6788 | D Band 1 #7 boundary MM | 1.56 |
| 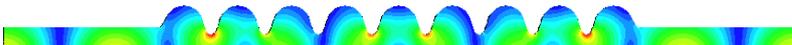 | 4.7736 | D Band 1 #8 boundary MM | 26.92 |
| 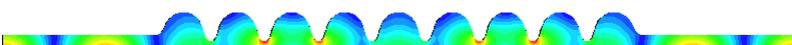 | 4.8443 | D Band 1 #9 boundary MM | 31.97 |
| | | | |
| 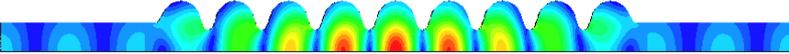 | 5.3518 | D Band 2 #1 boundary MM | 0.05 |
| 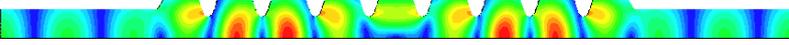 | 5.3927 | D Band 2 #2 boundary MM | 2.05 |
| 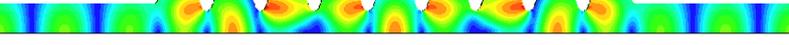 | 5.4279 | D Band 2 #3 boundary MM | 10.48 |
| 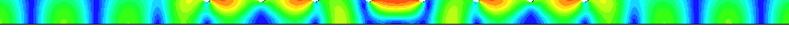 | 5.4540 | D Band 2 #4 boundary MM | 17.01 |
| 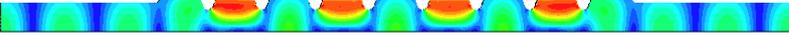 | 5.4727 | D Band 2 #5 boundary MM | 9.83 |
| 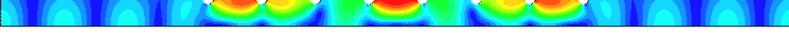 | 5.4854 | D Band 2 #6 boundary MM | 0.51 |
| 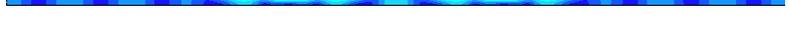 | 5.4930 | D Band 2 #7 boundary MM | 0.33 |
| 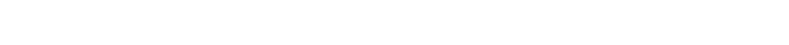 | 5.4969 | D Band 2 #8 boundary MM | 0.04 |
| 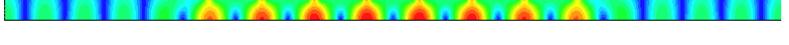 | 6.7962 | D Band 3 #1 boundary MM | 0.07 |
| 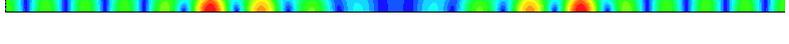 | 6.8243 | D Band 3 #2 boundary MM | 0.07 |
| 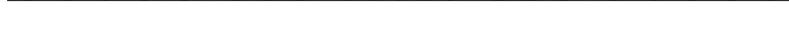 | 6.8913 | D Band 3 #3 boundary MM | 0.13 |
| 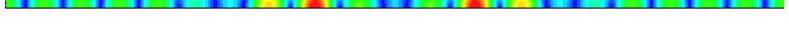 | 6.9888 | D Band 3 #4 boundary MM | 0.12 |
| 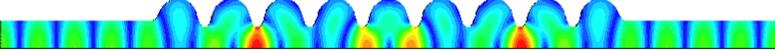 | 7.1000 | D Band 3 #5 boundary MM | 0.11 |
| 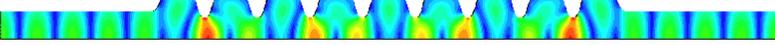 | 7.2154 | D Band 3 #6 boundary MM | 0.02 |
| 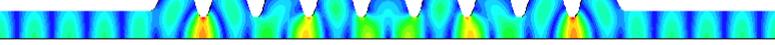 | 7.3369 | D Band 3 #7 boundary MM | 0.82 |
| 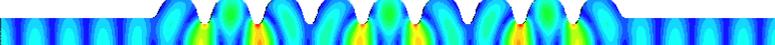 | 7.4629 | D Band 3 #8 boundary MM | 0.51 |
| 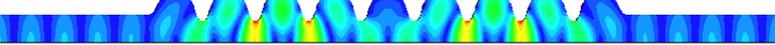 | 7.5787 | D Band 3 #9 boundary MM | 2.75 |
| | | | |
| 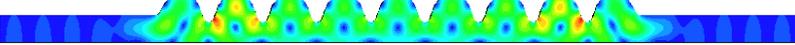 | 8.4971 | D Band 4 #1 boundary MM | 0.38 |
| 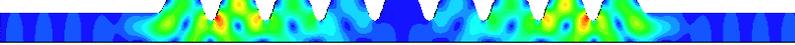 | 8.5008 | D Band 4 #2 boundary MM | 0.02 |
| 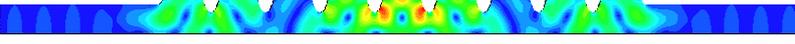 | 8.5288 | D Band 4 #3 boundary MM | 0.38 |
| 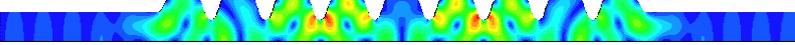 | 8.5707 | D Band 4 #4 boundary MM | 0.02 |
| 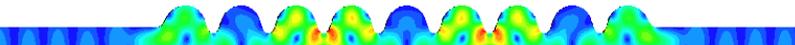 | 8.6275 | D Band 4 #5 boundary MM | 1.43 |
| 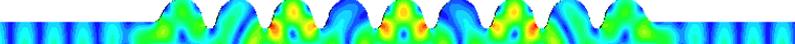 | 8.6861 | D Band 4 #6 boundary MM | 0.05 |
| 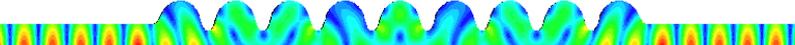 | 8.7279 | D Band 4 #7 boundary MM | 4.47 |
| 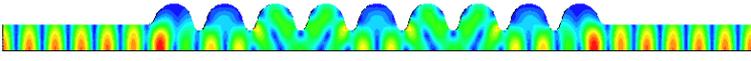 | 8.7719 | D Band 4 #8 boundary MM | 1.69 |
| 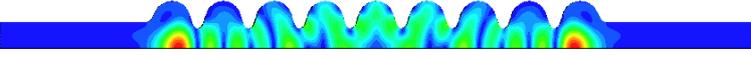 | 8.9182 | D Band 4 #9 boundary MM | 0.35 |
| | | | |
| 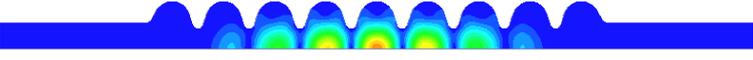 | 9.0563 | D Band 5 #1 boundary MM | 0.00 |
| 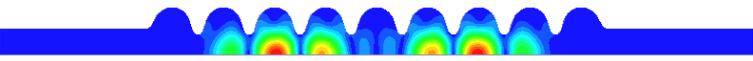 | 9.0571 | D Band 5 #2 boundary MM | 0.05 |
| 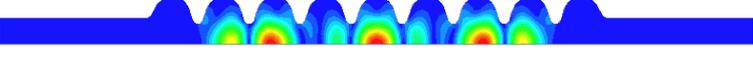 | 9.0588 | D Band 5 #3 boundary MM | 0.08 |
| 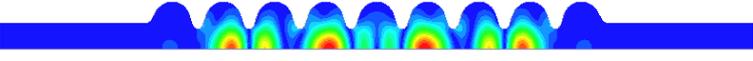 | 9.0624 | D Band 5 #4 boundary MM | 2.34 |
| 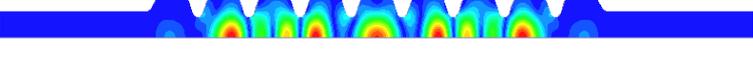 | 9.0710 | D Band 5 #5 boundary MM | 3.92 |
| 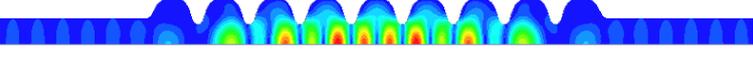 | 9.0950 | D Band 5 #6 boundary MM | 0.36 |
| | | | |
| 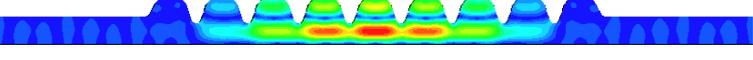 | 9.6955 | D Band 6 #1 boundary MM | 0.01 |
| 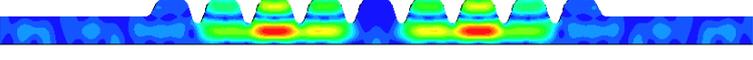 | 9.7162 | D Band 6 #2 boundary MM | 0.03 |
| 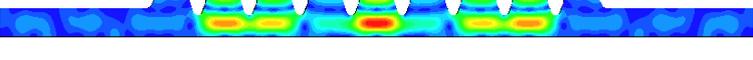 | 9.7476 | D Band 6 #3 boundary MM | 0.33 |
| 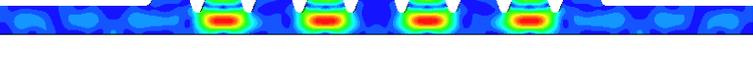 | 9.7840 | D Band 6 #4 boundary MM | 1.15 |
| 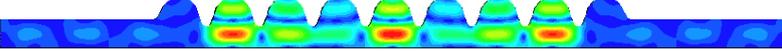 | 9.8201 | D Band 6 #5 boundary MM | 0.37 |
| 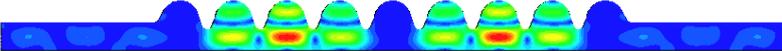 | 9.8510 | D Band 6 #6 boundary MM | 0.01 |
| 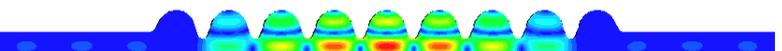 | 9.8721 | D Band 6 #7 boundary MM | 0.02 |

**Dipole cut-off parameters**

$TM_{01}$: $f_c$ = 5.7375 GHz for iris radius 15 mm
      $f_c$ = 7.6501 GHz for beam pipe radius 20 mm

$TE_{11}$: $f_c$ = 4.3920 GHz for iris radius 15 mm
      $f_c$ = 5.8560 GHz for beam pipe radius 20 mm

### Quadrupole modes

These quadrupole results were generated using HFSSv11 in which only a 90 degree slice of the structure was simulated using magnetic (M) symmetry planes. Magnetic (M) boundary conditions were applied to the ends of the beampipes. The meshing process consisted of: a volume mesh of applying a volumetric mesh restricted to a cell length no greater that 5mm; a similar mesh was applied on the surface of the structure in which the maximum mesh cell length was restricted to 5mm; in order to better account for the curved nature of the surface geometry a segmentation of 20 slices was applied. All simulations were run until a convergence of less than 0.05% was achieved. In the R/Q calculation in order to reduce the computational time required we considered only the electric energy in the calculation.

| Electric field pattern | $\omega/2\pi$ (GHz) | Band type | R/Q: $\Omega/cm^4$ |
|---|---|---|---|
| 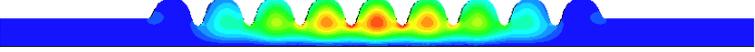 | 6.5620 | Q Band 1 #1 boundary MM | 0.18 |
| 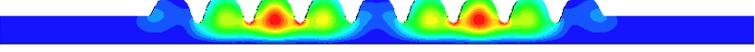 | 6.5828 | Q Band 1 #2 boundary MM | 3.75 |
| 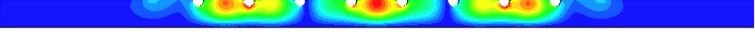 | 6.6154 | Q Band 1 #3 boundary MM | 4.38 |
| 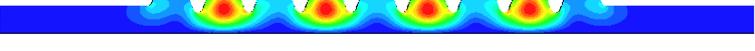 | 6.6574 | Q Band 1 #4 boundary MM | 0.19 |
| 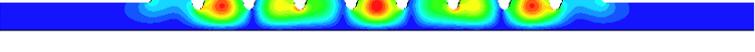 | 6.7043 | Q Band 1 #5 boundary MM | 0.15 |
| | | | |
| 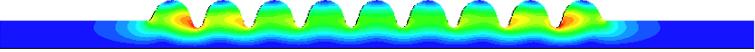 | 6.9974 | Q Band 2 #1 boundary MM | 0.13 |
| 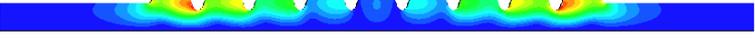 | 7.0065 | Q Band 2 #2 boundary MM | 0.08 |
| 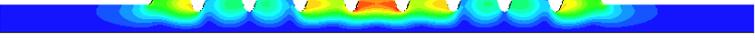 | 7.0400 | Q Band 2 #3 boundary MM | 0.15 |
| 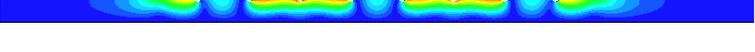 | 7.0797 | Q Band 2 #4 boundary MM | 0.00 |
| | | | |
| 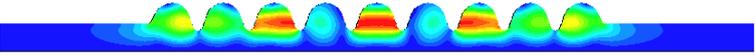 | 7.1132 | Q Band 2 #5 boundary MM | 0.22 |
| 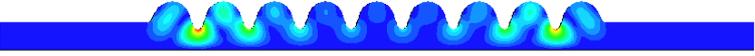 | 9.1127 | Q Band 3 #1 boundary MM | 3.66 |
| 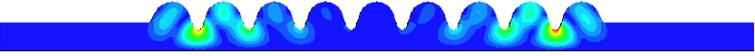 | 9.1134 | Q Band 3 #2 boundary MM | 0.11 |
| 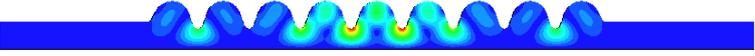 | 9.1214 | Q Band 3 #3 boundary MM | 6.98 |
| 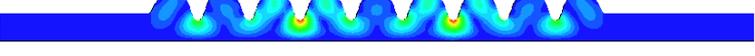 | 9.1324 | Q Band 3 #4 boundary MM | 11.28 |
| 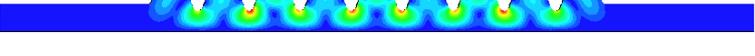 | 9.1477 | Q Band 3 #5 boundary MM | 2.75 |
| 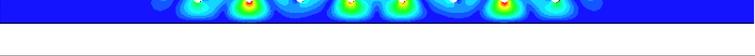 | 9.1663 | Q Band 3 #6 boundary MM | 0.01 |
| 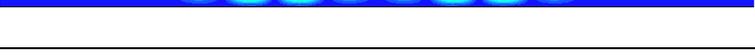 | 9.1858 | Q Band 3 #7 boundary MM | 0.15 |

**Quadrupole cut-off parameters**

$TM_{01}$: $f_c$ = 5.7375 GHz for iris radius 15 mm
$f_c$ = 7.6501 GHz for beam pipe radius 20 mm

$TE_{11}$: $f_c$ = 4.3920 GHz for iris radius 15 mm
$f_c$ = 5.8560 GHz for beam pipe radius 20 mm

## Sextupole modes

These sextupole results were generated using HFSSv11 in which only a 90 degree MM slice of the structure was simulated using both magnetic (M) and electric (E) symmetry planes. Magnetic (M) boundary conditions were applied to the ends of the beampipes. The meshing process consisted of: a volume mesh of applying a volumetric mesh restricted to a cell length no greater that 5mm; a similar mesh was applied on the surface of the structure in which the maximum mesh cell length was restricted to 5mm; in order to better account for the curved nature of the surface geometry a segmentation of 20 slices was applied. All simulations were run until a convergence of less than 0.05% was achieved. In the R/Q calculation in order to reduce the computational time required we considered only the electric energy in the calculation.

Here we have only considered the first sextupole band as we were interested in the band structure up to 10GHz, the majority of the second sextupole band lies above this frequency – refer to dispersion curve for a middle cell structure in the Dispersion curve section.

| Electric field pattern | $\omega/2\pi$ (GHz) | Band type | R/Q: $\Omega/cm^6$ |
|---|---|---|---|
| 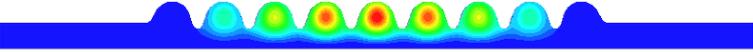 | 8.1890 | S Band 1 #1 boundary MM | 0.53 |
| 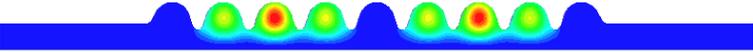 | 8.1936 | S Band 1 #2 boundary MM | 0.21 |
| 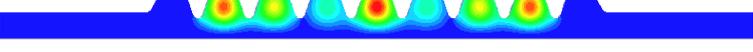 | 8.2005 | S Band 1 #3 boundary MM | 0.06 |
| 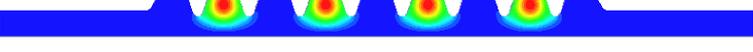 | 8.2090 | S Band 1 #4 boundary MM | 0.03 |
| 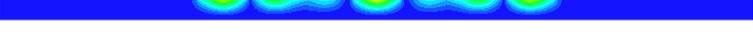 | 8.2175 | S Band 1 #5 boundary MM | 0.00 |
| 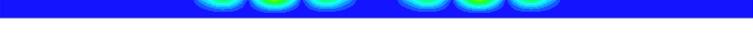 | 8.2248 | S Band 1 #6 boundary MM | 0.00 |
| 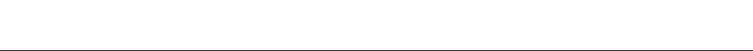 | 8.2300 | S Band 1 #7 boundary MM | 0.00 |
| 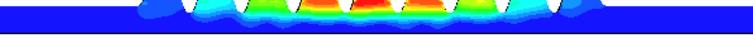 | 8.8018 | S Band 2 #1 boundary MM | 0.06 |
| 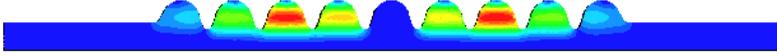 | 8.8086 | S Band 2 #2 boundary MM | 0.00 |
| 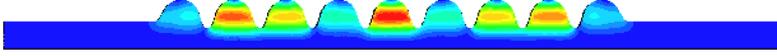 | 8.8178 | S Band 2 #3 boundary MM | 0.00 |
| 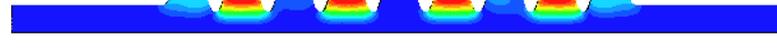 | 8.8275 | S Band 2 #4 boundary MM | 0.00 |
| 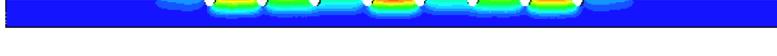 | 8.8373 | S Band 2 #5 boundary MM | 0.03 |
| 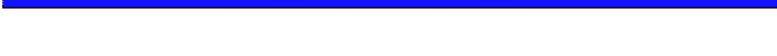 | 8.8452 | S Band 2 #6 boundary MM | 0.05 |
| 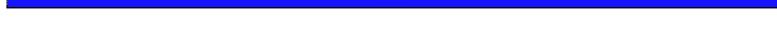 | 8.8501 | S Band 1 #7 boundary MM | 0.01 |

**Sextupole cut-off parameters**

$TM_{01}$: $f_c$ = 5.7375 GHz for iris radius 15 mm
    $f_c$ = 7.6501 GHz for iris radius 20 mm
$TE_{11}$: $f_c$ = 4.3920 GHz for iris radius 15 mm
    $f_c$ = 5.8560 GHz for iris radius 20 mm

# Comparison between the different boundary conditions

By modifying the boundary conditions used in the eigen mode simulations (from either EE or MM boundary conditions applied to the beam-pipes) presented in the previous sections, it is possible to observe whether a mode is trapped or multi-cavity in nature. Below is a direct comparison of this applied to the different bands.

**Monopole**

| Band | HFSS - EE | | | HFSS - MM | |
|---|---|---|---|---|---|
| | f: GHz | R/Q: Ω | | f: GHz | R/Q: Ω |
| 1 | 3.7483 | 0.009 | | 3.7483 | 0.009 |
| 1 | 3.7617 | 0.054 | | 3.7617 | 0.054 |
| 1 | 3.7825 | 0.075 | | 3.7825 | 0.075 |
| 1 | 3.8081 | 0.160 | | 3.8081 | 0.160 |
| 1 | 3.8357 | 0.294 | | 3.8357 | 0.294 |
| 1 | 3.8617 | 0.265 | | 3.8617 | 0.265 |
| 1 | 3.8832 | 0.153 | | 3.8832 | 0.153 |
| 1 | 3.8972 | 3.303 | | 3.8972 | 3.303 |
| 1 | 3.9028 | 372.705 | | 3.9028 | 372.705 |
| | | | | | |
| 2 | 7.043494 | 0.142 | | 7.0391 | 0.006 |
| 2 | 7.081938 | 0.976 | | 7.0685 | 0.034 |
| 2 | 7.138101 | 0.093 | | 7.1088 | 0.350 |
| 2 | 7.207645 | 2.924 | | 7.1535 | 0.212 |
| 2 | 7.283561 | 0.645 | | 7.2006 | 1.235 |
| 2 | 7.362203 | 2.673 | | 7.2581 | 3.400 |
| 2 | 7.437698 | 10.705 | | 7.3292 | 0.058 |
| 2 | 7.509145 | 19.122 | | 7.4105 | 15.076 |
| 2 | 7.582914 | 47.534 | | 7.4957 | 28.128 |
| | | | | | |
| 3 | 7.641745 | 0.594 | | 7.5768 | 3.604 |
| 3 | 7.723047 | 21.894 | | 7.6760 | 55.521 |
| 3 | 7.802748 | 0.229 | | 7.7562 | 0.109 |
| 3 | 7.879784 | 1.850 | | 7.8315 | 0.735 |
| 3 | 7.953646 | 0.639 | | 7.9007 | 0.004 |
| 3 | 8.021515 | 0.015 | | 7.9616 | 0.259 |
| 3 | 8.081535 | 0.113 | | 8.0142 | 0.693 |
| 3 | 8.129549 | 0.010 | | 8.0603 | 0.115 |
| 3 | 8.161176 | 0.002 | | 8.1023 | 0.731 |

**Dipole**

| Band | HFSS - EE | | | HFSS - MM | |
|---|---|---|---|---|---|
| | f: GHz | R/Q:Ω/cm$^2$ | | f: GHz | R/Q:Ω/cm$^2$ |
| 1 | 4.2953 | 0.00 | | 4.2951 | 0.00 |
| 1 | 4.3580 | 0.29 | | 4.3566 | 0.26 |
| 1 | 4.4460 | 0.00 | | 4.4285 | 0.07 |
| 1 | 4.5388 | 1.08 | | 4.4498 | 0.00 |
| 1 | 4.5972 | 0.79 | | 4.4751 | 0.32 |
| 1 | 4.6399 | 0.16 | | 4.5683 | 1.21 |
| 1 | 4.7227 | 10.37 | | 4.6788 | 1.56 |
| 1 | 4.8312 | 50.20 | | 4.7736 | 26.92 |
| 1 | 4.9260 | 30.38 | | 4.8443 | 31.97 |
| | | | | | |
| 2 | 5.3583 | 0.04 | | 5.3518 | 0.05 |
| 2 | 5.4058 | 5.01 | | 5.3927 | 2.05 |
| 2 | 5.4441 | 20.88 | | 5.4279 | 10.48 |
| 2 | 5.4696 | 16.07 | | 5.4540 | 17.01 |
| 2 | 5.4849 | 0.98 | | 5.4727 | 9.83 |
| 2 | 5.4933 | 1.25 | | 5.4854 | 0.51 |
| 2 | 5.4973 | 0.31 | | 5.4930 | 0.33 |
| 2 | 5.4982 | 0.48 | | 5.4969 | 0.04 |
| | | | | | |
| 3 | 6.8225 | 0.01 | | 6.7962 | 0.07 |
| 3 | 6.8994 | 0.03 | | 6.8243 | 0.07 |
| 3 | 7.0022 | 0.06 | | 6.8913 | 0.13 |
| 3 | 7.1227 | 0.18 | | 6.9888 | 0.12 |
| 3 | 7.2551 | 0.55 | | 7.1000 | 0.11 |
| 3 | 7.3852 | 0.01 | | 7.2154 | 0.02 |
| 3 | 7.4915 | 0.45 | | 7.3369 | 0.82 |
| 3 | 7.5654 | 0.27 | | 7.4629 | 0.51 |
| 3 | 7.6235 | 1.27 | | 7.5787 | 2.75 |
| | | | | | |
| 4 | 8.4995 | 0.13 | | 8.4971 | 0.38 |
| 4 | 8.5045 | 0.09 | | 8.5008 | 0.02 |
| 4 | 8.5318 | 0.16 | | 8.5288 | 0.38 |
| 4 | 8.5762 | 0.11 | | 8.5707 | 0.02 |
| 4 | 8.6399 | 0.42 | | 8.6275 | 1.43 |
| 4 | 8.7207 | 1.04 | | 8.6861 | 0.05 |
| 4 | 8.7998 | 10.06 | | 8.7279 | 4.47 |
| 4 | 8.8613 | 2.57 | | 8.7719 | 1.69 |
| 4 | 8.9167 | 0.39 | | 8.9182 | 0.35 |
| | | | | | |
| 5 | 9.0560 | 0.00 | | 9.0563 | 0.00 |
| 5 | 9.0568 | 0.05 | | 9.0571 | 0.05 |
| 5 | 9.0585 | 0.07 | | 9.0588 | 0.08 |
| 5 | 9.0620 | 2.17 | | 9.0624 | 2.34 |
| 5 | 9.0703 | 4.04 | | 9.0710 | 3.92 |
| 5 | 9.0933 | 0.55 | | 9.0950 | 0.36 |

### **Quadrupole**

| Band | HFSS - EE | | | HFSS - MM | |
|---|---|---|---|---|---|
| | f: GHz | R/Q:$\Omega/cm^4$ | | f: GHz | R/Q:$\Omega/cm^4$ |
| 1 | 6.5625 | 0.18 | | 6.5620 | 0.18 |
| 1 | 6.5832 | 3.75 | | 6.5828 | 3.75 |
| 1 | 6.6158 | 4.38 | | 6.6154 | 4.38 |
| 1 | 6.6578 | 0.19 | | 6.6574 | 0.19 |
| 1 | 6.7057 | 0.31 | | 6.7043 | 0.15 |
| | | | | | |
| 2 | 6.9977 | 0.14 | | 6.9974 | 0.13 |
| 2 | 7.0067 | 0.08 | | 7.0065 | 0.08 |
| 2 | 7.0429 | 0.15 | | 7.0400 | 0.15 |
| 2 | 7.0798 | 0.00 | | 7.0797 | 0.00 |
| 2 | 7.1133 | 0.22 | | 7.1132 | 0.22 |
| | | | | | |
| 3 | 9.1103 | 1.86 | | 9.1127 | 3.66 |
| 3 | 9.1106 | 0.01 | | 9.1134 | 0.11 |
| 3 | 9.1204 | 2.39 | | 9.1214 | 6.98 |
| 3 | 9.1314 | 4.74 | | 9.1324 | 11.28 |
| 3 | 9.1468 | 2.26 | | 9.1477 | 2.75 |
| 3 | 9.1656 | 0.00 | | 9.1663 | 0.01 |
| 3 | 9.1853 | 0.08 | | 9.1858 | 0.15 |

### **Sextupole**

| Band | HFSS - EE | | | HFSS - MM | |
|---|---|---|---|---|---|
| | f: GHz | R/Q:$\Omega/cm^4$ | | f: GHz | R/Q:$\Omega/cm^4$ |
| 1 | 8.1890 | 0.53 | | 8.1890 | 0.53 |
| 1 | 8.1936 | 0.21 | | 8.1936 | 0.21 |
| 1 | 8.2005 | 0.06 | | 8.2005 | 0.06 |
| 1 | 8.2090 | 0.03 | | 8.2090 | 0.03 |
| 1 | 8.2175 | 0.00 | | 8.2175 | 0.00 |
| 1 | 8.2248 | 0.00 | | 8.2248 | 0.00 |
| 1 | 8.2300 | 0.00 | | 8.2300 | 0.00 |
| | | | | | |
| 2 | 8.8018 | 0.14 | | 8.8018 | 0.14 |
| 2 | 8.8086 | 0.08 | | 8.8086 | 0.08 |
| 2 | 8.8178 | 0.15 | | 8.8178 | 0.15 |
| 2 | 8.8275 | 0.00 | | 8.8275 | 0.00 |
| 2 | 8.8373 | 0.22 | | 8.8373 | 0.22 |
| 2 | 8.8452 | 0.05 | | 8.8452 | 0.05 |
| 2 | 8.8501 | 0.01 | | 8.8501 | 0.01 |

# Dispersion curves

The dispersion curves for the 3.9GHz bunchshaping cavity structure [1], together with synchronous points, up to the second sextupole i.e. below 11GHz are displayed below in figure 1  All simulations were made using HFSSv11 until a convergence below 0.005% was achieved; the meshing and symmetry planes used to obtain the various bands are the same as those outlined in the pictorial mode distribution above.

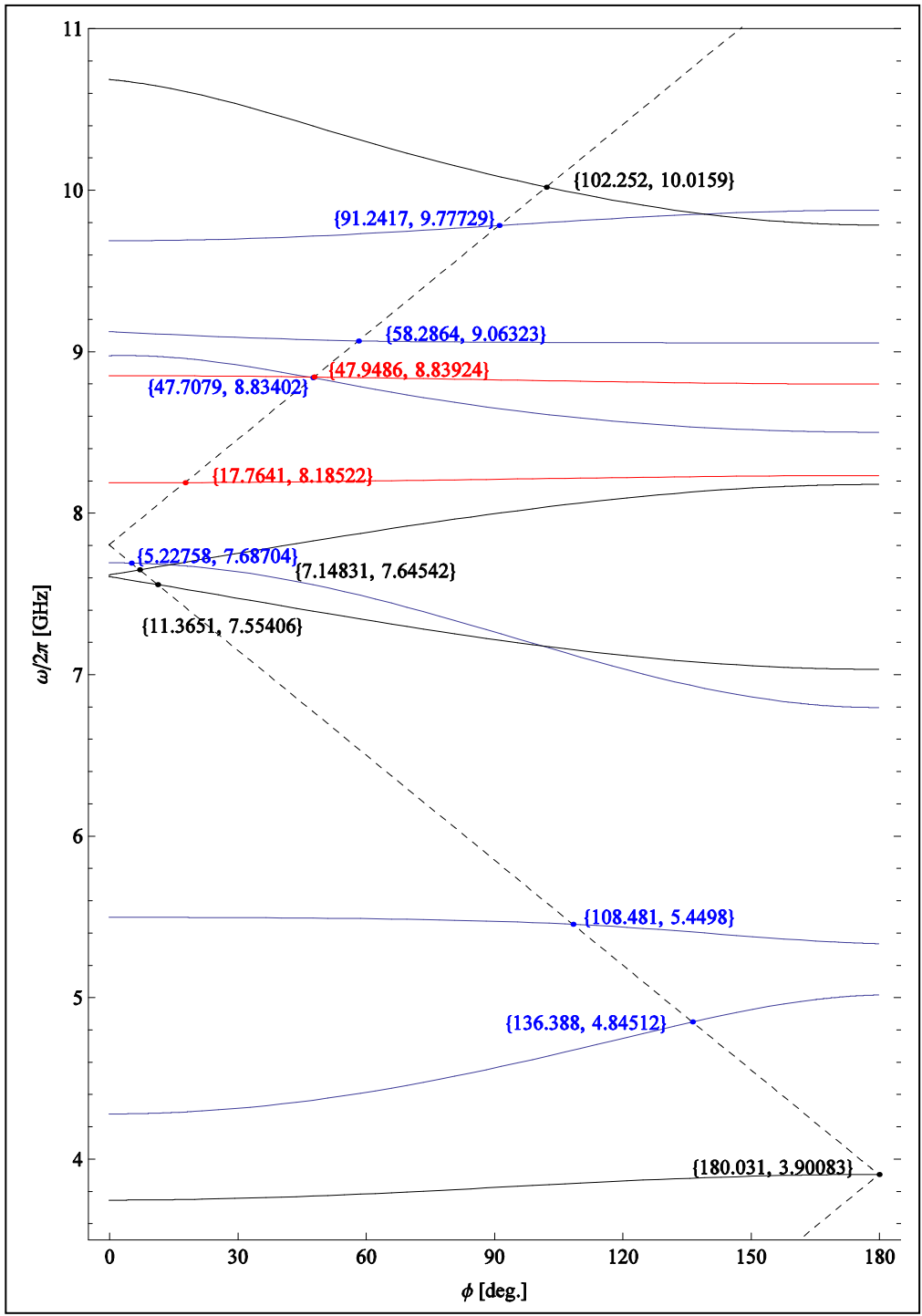

Figure 1. Single cell dispersion curves with synchronous points; monopole (black), dipole (blue), sextupole (red).

# Circuit models

## Circuit model of monopole modes

A study involving the circuit model applied to the monopole bands was conducted in [3], in which it was shown that the traditional nearest neighbour coupling model ($n_c$=2) found in the literature ( i.e. by Nagle, Knapp and Knapp [4] ) is a good representation for the first band; however there is a significant discrepancy between this circuit model and that numerically predicted by HFSS for higher bands as seen below in figure 2a Adding coupling from two additional neighbours ($n_c$=4), as derived in [3], improves the prediction significantly and this is displayed in figure 2b

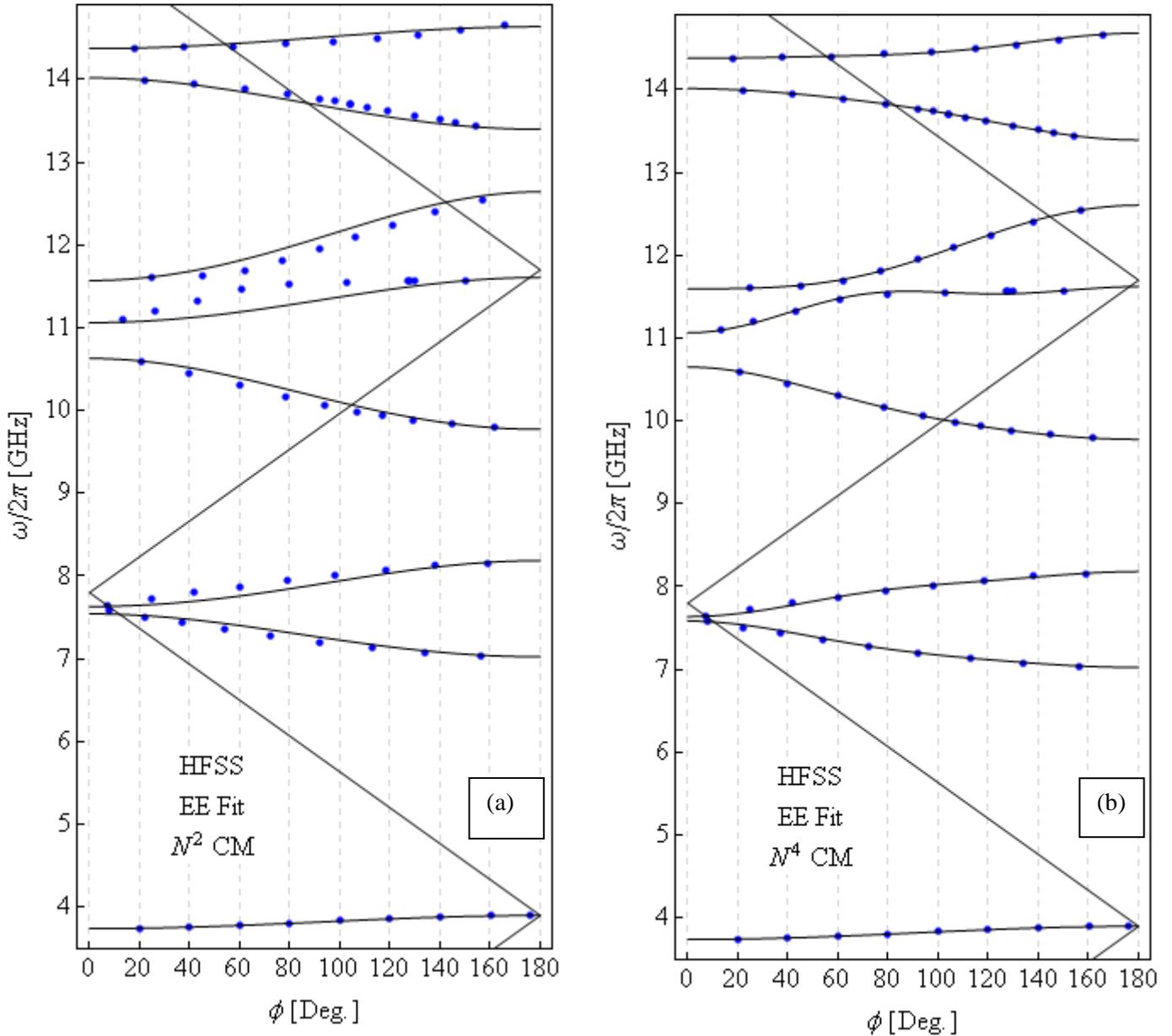

Figure 2: 9-cell simulations from HFSS v11 with MM planes (blue) and analytical dispersion curves from a) the classical model ($n_c$=2) (black) and b) the extended model ($n_c$=4) (black).

## Circuit model of dipole modes

Here we provide an initial investigation into using the double chain circuit model developed by Bane and Gluckstern [5], to the 3.9GHz bunch shaping cavity. This circuit model gives a fairly good agreement to that predicted by the numerical simulation, as can be seen in the dispersion relation of the first two dipole bands in figure 3 below.

This nearest neighbour double chain circuit model is also good at representing the simulation R/Q's of the dipole modes for the first two dipole bands, refer to figure 4. As a first means of rapid analysis, the model in the literature is adequate, i.e. the main advantage of the circuit model is that one can quickly obtain the band structure and R/Q's in a matter of hours as compared to a full day of numerical simulation. The analogy we propose here is similar to the case for the monopole circuit model, the addition of further cell coupling will improve these results and those predicted for the higher dipole bands; this remains an aspect for further future development.

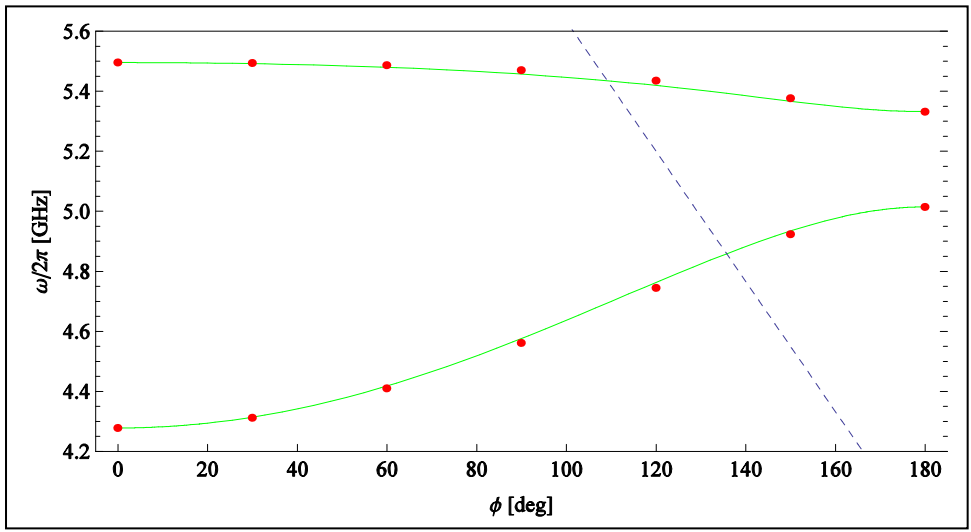

Figure 3. Dispersion curve of first two dipole band, red dots are for the single cell simulation from HFSS, green curves are the double chain circuit model prediction.

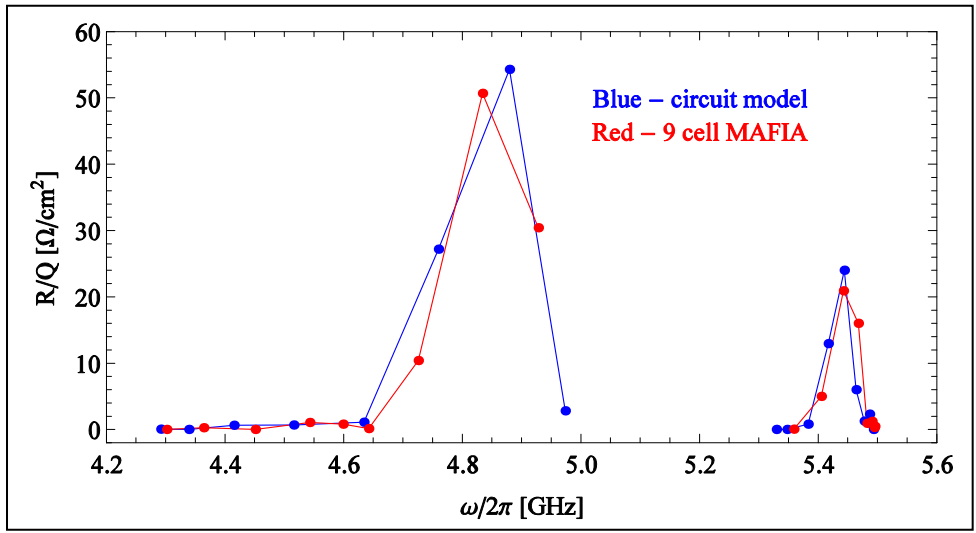

Figure 4. Dipole R/Q comparison between circuit model results and MAFIA simulation results

# References

[1] T. Khanibouline, N. Solyak, R. Wanzenberg, HOMs of a 3rd harmonic cavity with increased end-cup iris, FERMILAB-TM-2210, 2003.
[2] R. Wanzenberg, Monopole, Dipole and Quadrupole Passbands of the TESLA 9-cell Cavity, TESLA 2001-33, 2001.
[3] B. Szczesny, I.R.R.Shinton and R.M.Jones, Third Harmonic Cavity Modal Analysis, 14th International Conference on RF Superconductivity, Berlin, Germany, 20-25 Sep 2009.
[4] D.E. Nagle, E.A. Knapp, B.C. Knapp, Coupled Resonator Model of Standing Wave Accelerator Tanks, Rev.Sci.Instr. **39**, 11, 1583-1587, 1967.
[5] K.L. Bane and R.L. Gluckstern, The Transverse Wakefield of a Detuned X-Band Accelerating Structure, Part. Accel. **42**, 123-169, 1993.
[6] N. Juntong, Roger M. Jones, I.R.R. Shinton, C.D. Beard, G. Burt, Proceedings of 11th European Particle Accelerator Conference (EPAC 08), Magazzini del Cotone, Genoa, Italy, 23-27 Jun 2008.